\documentclass[11pt,a4paper]{article}
\pdfoutput=1
\usepackage{jheppub}
\usepackage[utf8]{inputenc}
\usepackage[T1]{fontenc}
\usepackage[english]{babel}
\usepackage{graphicx}
\usepackage{amsmath}
\usepackage{color}
\usepackage{physics}
\usepackage{listings}
\usepackage{hyperref}
\usepackage{orcidlink}
\usepackage{todonotes}
\usepackage{subfigure}

\title{From Hubble to Bubble}

\author[1]{Maciej~Kierkla \orcidlink{0000-0002-2785-5370},}
\author[2]{Giorgio~Laverda \orcidlink{0000-0002-4739-4946},}
\author[1]{Marek~Lewicki \orcidlink{0000-0002-8378-0107},}
\author[1]{Andreas~Mantziris \orcidlink{0000-0001-5899-670X},}
\author[2]{Matteo~Piani \orcidlink{0000-0002-2387-5948},}
\author[3]{Javier~Rubio \orcidlink{0000-0001-7545-1533}}
\author[1]{and Mateusz~Zych \orcidlink{0000-0003-3675-5037}}
\affiliation[1]{Faculty of Physics, University of Warsaw, ul.\ Pasteura 5, 02-093 Warsaw, Poland}
\affiliation[2]{Centro de Astrofísica e Gravitação - CENTRA, Departamento de Física, Instituto Superior
Técnico - IST, Universidade de Lisboa - UL, Av. Rovisco Pais 1, 1049-001 Lisboa, Portugal}
\affiliation[3]{Departamento de Física Teórica and Instituto de Física de Partículas y del Cosmos (IPARCOS-UCM), Universidad Complutense de Madrid, 28040 Madrid, Spain}
\emailAdd{maciej.kierkla@fuw.edu.pl}
\emailAdd{giorgio.laverda@tecnico.ulisboa.pt}
\emailAdd{marek.lewicki@fuw.edu.pl}
\emailAdd{andreas.mantziris@fuw.edu.pl}
\emailAdd{matteo.piani@tecnico.ulisboa.pt}
\emailAdd{javier.rubio@ucm.es}
\emailAdd{mateusz.zych@fuw.edu.pl}
\abstract{
The detection of a stochastic Gravitational Wave (GW) background sourced by a cosmological phase transition would allow us to see the early Universe from a completely new perspective, illuminating aspects of Beyond the Standard Model (BSM) physics and inflationary cosmology.
In this study, we investigate whether the evolution of the scalar potential of a minimal SM extension after inflation can lead to a strong first-order phase transition. In particular, we focus on a BSM spectator scalar field that is non-minimally coupled to gravity and has a dynamical double-well potential. As inflation ends, the potential barrier diminishes due to the evolution of the curvature scalar. Therefore, a phase transition can proceed through the nucleation of true-vacuum bubbles that collide as they fill the Universe and produce GWs. We consider high and low scales of inflation, while also taking into account a kination period between inflation and the onset of radiation domination. With this prescription, we showcase a proof-of-concept study of a new triggering mechanism for BSM phase transitions in the early Universe, whose GW signatures could potentially be probed with future detectors. 
}

\begin{document}
\maketitle

%%%%%%%%%%%%%%%%%%%%%%%%%%%%%%%%%%%%%%%%%%%%%%%%%%%%%%%%%%%%%%%%%%%%%%%%
\section{Introduction}
%%%%%%%%%%%%%%%%%%%%%%%%%%%%%%%%%%%%%%%%%%%%%%%%%%%%%%%%%%%%%%%%%%%%%%%%

There exists a long-lasting link between first-order phase transitions and the very early history of the Universe. This relationship stretches from the first works on inflation of the early 80s~\cite{Guth:1980zm, Linde:1981zj, Mukhanov:1981xt} up to the modern interpretation as beyond-the-standard-model phenomenology. Indeed, nowadays the most compelling settings for such transitions are realised at the quark-hadron transition scale~\cite{Iso:2017uuu, Sagunski:2023ynd} and the electroweak scale (EW)~\cite{Caprini:2015zlo, Caprini:2019egz}. Given the thermalised state of the Universe at both energy scales, the evolution of true-vacuum bubbles proceeds with a non-trivial interplay between the scalar sector and the primordial plasma~\cite{LISACosmologyWorkingGroup:2022jok}. The end results can vary and constitute the main motivation to investigate first-order phase transitions: from the generation of baryon asymmetry~\cite{Morrissey:2012db, Cline:2020jre, Cline:2021dkf, Lewicki:2021pgr, Ellis:2022lft} through the production of a stochastic background of gravitational waves~\cite{Caprini:2015zlo, Caprini:2019egz, AEDGE:2019nxb, Badurina:2021rgt} to the seeding of primordial magnetic fields~\cite{Vachaspati:1991nm, Brandenburg:2017neh, Ellis:2019tjf, Ellis:2020uid, RoperPol:2023bqa} and primordial black holes~\cite{Hawking:1982ga, Kodama:1982sf, Lewicki:2019gmv, Kawana:2021tde, Liu:2021svg, Hashino:2022tcs, Huang:2022him, Kawana:2022lba, Kawana:2022olo, Lewicki:2023ioy, Gouttenoire:2023naa, Gouttenoire:2023bqy}. The standard picture of a thermal phase transition \cite{Linde:1980tt, Linde:1981zj} entails the nucleation of bubbles, their collision, the interaction between bubbles and plasma, and the final percolation in a thermalised medium. The potential barrier responsible for the separation between vacua is typically generated by thermal contributions to the scalar effective potential and tunnelling becomes efficient as the Universe cools down to the energy scale of the process.

In this work, we present an innovative and flexible mechanism beyond the traditional thermal picture to trigger a First-Order Phase Transition (FOPT), where the formation of true-vacuum bubbles is simply induced by the evolution of the Hubble rate in the primordial Universe. The field content of the model is kept to a minimum by considering only the inflaton and a subdominant spectator scalar field. The inflaton is responsible for the overall background cosmological dynamics during and immediately after inflation. The spectator field is a prototypical component of a beyond-the-standard-model sector, and it is characterised by a direct coupling to the background curvature. At the heart of the mechanism is precisely the non-minimal interaction between quantum fields and the geometry of spacetime at high energies, which constitutes a simple cosmic clock that sets off the phase transition soon after inflation (see for instance Refs. \cite{Bettoni:2019dcw, Dimopoulos:2001ix, Bassett:1997az}). Indeed, at the end of the slow-roll phase, the overall cosmological equation of state (e.o.s.) evolves from $w=-1$ to $w>-1$ since the inflationary kinetic energy density becomes dominant. The spectator field experiences the change of e.o.s. as a variation of its time-dependent effective mass. A first-order vacuum phase transition occurs if the decreasing effective mass becomes comparable to the negative cubic self-coupling of the spectator field. Bubbles of true vacuum form via tunnelling through the potential barrier, and in turn, gravitational waves are produced by bubble-wall collisions.   

This novel mechanism is minimal and natural, as it relies only on the presence of a Hubble-dependent effective mass of the spectator field. Its simplicity allows the mechanism to be easily included in the phenomenology of the early Universe independently of the specific choice of inflationary scenario. This feature makes it a flexible tool for studying the production of stochastic backgrounds of gravitational waves in a wide range of frequencies, where the spectrum’s peak frequency is determined by the energy scale at the time of the transition.

For simplicity, in this work we focus on the prototypical case of a spectator field non-minimally coupled to the Ricci curvature, a setup analogous to Hubble-induced second-order phase transitions \cite{Bettoni:2021zhq, Laverda:2023uqv, Bettoni:2019dcw} and related scenarios \cite{Figueroa:2016dsc,Opferkuch:2019zbd, Dimopoulos:2018wfg,Bettoni:2018utf,Bettoni:2018pbl,Babichev:2020xeg,Babichev:2020yeo,Freese_2022}. The spectator potential contains the usual renormalisable operators up to fourth order, a minimal choice that still endows all  the necessary features for the desired phenomenology. The background dynamics are defined by a quintessential inflation-like scenario with an epoch of kinetic domination ($w=1$) following the end of the slow-roll phase (see Ref. \cite{Bettoni:2021qfs} for a review). This choice comes with the benefits of realising one single first-order phase transition and of amplifying the gravitational waves signal throughout kination. We assume a specific sigmoid function for the evolution of the equation-of-state parameter $w(t)$ with one free parameter determining the speed of the shift from $w=-1$ to $w=1$. The nucleation rate and the energy released as bubbles can be computed from the O(4)-symmetric Euclidean action for the critical bubble~\cite{Coleman:1977py} and indicate that the transition happens rapidly, even before the Universe fully enters the kination epoch. We then estimate the typical size of the bubbles at percolation time, which ultimately allows us to study the characteristics of the gravitational wave spectrum produced by wall collisions. Its peak frequency and amplitude are completely determined by a few parameters: the Hubble scale at the end of inflation, the speed of the shift from inflation to kination, the typical bubble size at percolation and the strength of the transition proportional to the energy difference between the vacua. Finally, we discuss the implications of such curvature-induced phase transitions for the upcoming gravitational wave observatories.

This work is organised as follows. Section~\ref{sec:scalar_potential} introduces the evolution of the cosmological background and the non-minimal interaction of the BSM scalar sector with gravity. There, the spectator potential is defined alongside a parameterisation for the inflation-kination transition. Section~\ref{sec:bubbles} focuses on the nucleation process and the production of gravitational waves through bubble-wall collisions. Section~\ref{sec:gw} presents the results of scanning the available parameter space and shows the shape of the estimated gravitational-waves spectra at present time. In Section~\ref{sec:conclusions} we discuss our findings and summarise the results.

%%%%%%%%%%%%%%%%%%%%%%%%%%%%%%%%%%%%%%%%%%%%%%%%%%%%%%%%%%%%%%%%%%%%%%%%
\section{Cosmological evolution of the scalar potential} \label{sec:scalar_potential}
%%%%%%%%%%%%%%%%%%%%%%%%%%%%%%%%%%%%%%%%%%%%%%%%%%%%%%%%%%%%%%%%%%%%%%%%

First-order phase transitions are a common feature of many theories beyond the Standard Model~\cite{LISACosmologyWorkingGroup:2022jok}. In the context of cosmological phase transitions, the evolution of a scalar field is typically characterised by a change from an initial high-temperature phase in which the symmetry is restored to the global minimum of scalar potential. In the most common case, thermally driven transitions are considered; however, a FOPT can also proceed through quantum tunnelling, where the impact of temperature corrections is negligible as long as the scalar potential has all the necessary properties. To illustrate this type of transition we consider a minimal extension of the Standard Model, featuring an extra scalar singlet $\chi$, endowed with a renormalisable potential and a non-minimal coupling to gravity. In the action
\begin{equation}
    S=\int d^4 x \sqrt{-g} \left[\frac{M_P^2-\xi \chi^2}{2}\mathcal{R}-\frac{1}{2}\partial_\mu \chi \partial^\mu \chi- \frac{m^2}{2} \chi^2 + \frac{\sigma}{3}\chi^3-\frac{\lambda}{4}\chi^4 \right]\,,
    \label{eq:scalar-action}
\end{equation}
$M_P=2.435 \times 10^{18}$ GeV is the reduced Planck mass, $\xi$ the non-minimal coupling\footnote{As this action does not feature any $Z_2$ symmetry, one could introduce a second non-minimal coupling $g \phi \mathcal{R}$ for completeness. Since this operator affects only the position of the vacua, it will be neglected in what follows.}, $\mathcal{R}$ the Ricci scalar, $m$ the mass of the field in flat spacetime, and $\sigma$ and $\lambda$ the cubic and quartic self-couplings, respectively. All of the model parameters are taken to be real and positive. The motivation for the inclusion of the non-minimal coupling to curvature is two-fold. Firstly, its presence is necessary to ensure the renormalisability of the energy-momentum tensor in curved space-time \cite{Birrell:1982ix}. Secondly, such a term is fully compliant with the symmetries of the action, and it will arise from radiative corrections, even if it is set to zero at a specific scale. For a flat FLRW metric, the Ricci scalar is given by
\begin{equation}\label{eq:Ricci}
    \mathcal{R}=3(1-3w)H^2\,,
\end{equation}
where $w=p/\rho$ is the e.o.s. parameter relating the pressure and energy density of the cosmic fluid, according to the dominating component of the energy-momentum tensor. Assuming that, in the initial stage of the cosmic evolution, the field $\chi$ is subdominant and does not affect the dynamics of $w$, we can treat the non-minimal coupling as a contribution to the effective mass. Thus, the effective potential is given by
\begin{equation}
    \label{eq:eff-potential}
    V_{\rm eff}= \frac{M^2}{2}\chi^2-\frac{\sigma}{3}\chi^3+\frac{\lambda}{4}\chi^4\,, \hspace{10mm}    M^2= m^2+\xi \mathcal{R}\,.
\end{equation} 
It is clear from Eq.~\eqref{eq:Ricci} that an increasing e.o.s.~parameter corresponds to a decreasing effective mass, which becomes negative for fluids stiffer than radiation $w>1/3$. In the regime $0<M^2<2\sigma^2/(9\lambda)$, the effective potential in Eq.~\eqref{eq:eff-potential} has a false vacuum in $\chi=0$ and a true vacuum at 
\begin{equation}
    \label{eq:true-vacuum}
    \chi_{\rm tv}=\frac{\sigma}{2\lambda}\left(1+\sqrt{1-\frac{4 M^2 \lambda}{\sigma^2}}\right)\,.
\end{equation}

The existence of two different vacua sets the stage for a cosmological first-order phase transition in the post-inflationary epoch. During inflation, the inflaton field is slowly rolling down an almost flat potential with negligible kinetic energy. In this regime, the e.o.s parameter is $w_{\rm inf}=-1$, the Hubble parameter is approximately constant, and the Ricci scalar gives a positive contribution to the effective mass so that $M^2$ is maximal. As inflation ends, the inflaton leaves the plateau region of its potential and starts to roll down towards its minimum, until all potential energy is converted into kinetic and $w_{\rm kin}=1$. In this context, our study is performed under the following simplifying assumptions:
\begin{enumerate}
    \item The field $\chi$ remains energetically subdominant until the phase transition takes place.
    \item The mass of the field in flat spacetime is much smaller than the gravitational contribution and thus, we can safely set $m=0$.
    \item A period of kinetic domination (kination) follows the end of inflation.
\end{enumerate}

The first condition ensures that the field does not play any role in the initial dynamics of the e.o.s.~parameter $w(t)$, allowing for a model-independent parameterisation of its evolution. The second condition ensures that the potential always enters a broken phase until it eventually turns tachyonic around the origin ($\chi=0$) when $w=1/3$. In principle, one could consider a non-negligible value for $m$ but this would lead to a more complicated analysis due to the non-trivial interplay between $m$, $\xi$ and $H$. The third condition, albeit not strictly necessary, offers some conceptual advantages. For typical models of single-field inflation, the inflaton starts oscillating around its minimum at the end of the slow-roll phase and the e.o.s parameter will periodically change value between $-1\leq w \leq 1$. In this case, the potential of the spectator field would also periodically shift between phases of broken and restored symmetry. In doing so, a series of tunnelling and rolling events might ensue, but the study of their non-trivial dynamics goes beyond the scope of this work. Moreover, in the absence of any heating mechanism, the radiation produced during the phase transition will grow over the kination-dominated background, and therefore it will naturally fix the duration of the heating stage. In the following analysis, we will assume that the Universe reaches thermal equilibrium as soon as it enters the radiation domination (RD) epoch.

\begin{figure}[t]
    \centering
    \includegraphics[scale=1]{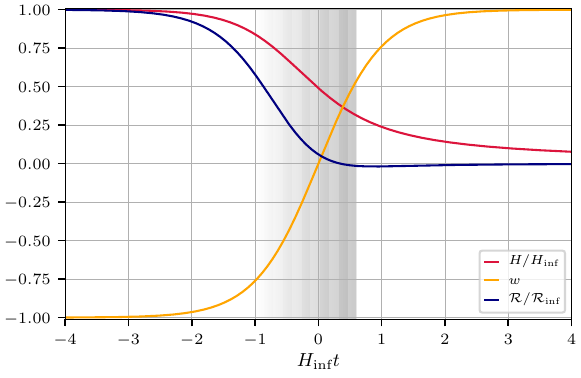}
    \caption{\it The evolution of the e.o.s.~parameter, the Hubble rate and the Ricci scalar from inflation to kination. 
    The rate of change is set to $\beta_w/H_{\rm inf} = 1$. The shaded area denotes the period in which a FOPT can naturally occur.}
    \label{fig:w-H-R}
\end{figure}

Parameterising the evolution of the global e.o.s.~parameter as
\begin{align}
    w(t)=\tanh{(\beta_w (t-t_{0}))} \, ,
    \label{eq:omega}
\end{align}
allows us to perform explicit analytical computations, where $t_0$ is the time at which $w=0$ and the free parameter $\beta_w > 0$ controls the speed of the transition between boundary values. The corresponding Hubble rate is given by
\begin{align}
    H(t)=\left[\frac32 \left(t + \frac{1}{\beta_w} \ln{\left[ \cosh{(\beta_w (t-t_{0}))} \right]} + c \right)\right]^{-1} \, ,
\end{align}
where $c=\frac{2}{3H_{\rm inf}} + \frac{\ln{2}}{\beta_w} - t_{0}$ is an integration constant and $H_{\rm inf} = H (t \rightarrow - \infty) $  is the scale of inflation. For convenience, we rewrite this expression in terms of Hubble time as
\begin{align}\label{eq:H(t)}
    \frac{H(t)}{H_{\rm inf}}= \frac{2}{3} \left(H_{\rm inf}(t-t_{0}) + \left( \frac{\beta_w}{H_{\rm inf}} \right)^{-1} \ln{\left[ 2 \cosh{\left( \left(\frac{\beta_w}{H_{\rm inf}} \right) H_{\rm inf} (t-t_{0})\right)} \right]} + \frac{2}{3} \right)^{-1}\, .
\end{align}
Figure \ref{fig:w-H-R} displays the time evolution of $H(t)$, $w(t)$ and $R(t)$ according to the chosen parameterisation in Eq.~\eqref{eq:omega}. 

%%%%%%%%%%%%%%%%%%%%%%%%%%%%%%%%%%%%%%%%%%%%%%%%%%%%%%%%%%%%%%%%%%%%%%%%
\section{Bubble nucleation and gravitational wave production from collisions} \label{sec:bubbles}
%%%%%%%%%%%%%%%%%%%%%%%%%%%%%%%%%%%%%%%%%%%%%%%%%%%%%%%%%%%%%%%%%%%%%%%%
In order to calculate the necessary parameters describing the vacuum transition one has to be able to fully describe the evolution of scalar potential and then calculate the tunnelling probability. For the scope of this work, we shall adopt a semi-analytical method described in \cite{Adams:1993zs}. The scalar potential in Eq.~\eqref{eq:eff-potential} can be expressed in the reduced dimensionless form as
\begin{equation}
    \widetilde{V}(\Tilde{\varphi}, t) = \frac{1}{4}\Tilde{\varphi}^4 - \Tilde{\varphi}^3 +\frac{\delta(t)}{2}\Tilde{\varphi}^2 \, ,
\end{equation}
where 
\begin{equation}
\label{eq:delta}
\Tilde{\varphi} = \frac{3\lambda}{\sigma}\chi \qquad \textrm{and}\qquad\delta(t) = \frac{9 M(t)^2\lambda}{\sigma^2} \, .
\end{equation}
The parameter $\delta$ fully controls the evolution of the potential and therefore, it determines when the transition takes place. Its value varies from $2$ to $0$, as it follows the evolution of the effective mass term from degenerate minima to a vanishing barrier, respectively. In this formalism, the position of the true vacuum minimum (\ref{eq:true-vacuum}) reads
\begin{align}
     \chi_{\rm tv} &= \frac{\sigma} {2 \lambda} \left( 1 + \sqrt{1- \frac{4 \delta}{9}} \right) \,.
\end{align}
Therefore, the energy difference $\Delta V$ between the minima can be expressed as
\begin{equation}
\begin{split}
    \Delta V= V(0) - V(\chi_{\rm tv})= \frac{\sigma^4}{96 \lambda^3} \left( 1 + \sqrt{1- \frac{4 \delta}{9}} \right)^2  \left( 1 - \frac{2 \delta}{3} + \sqrt{1- \frac{4 \delta}{9}} \right). 
\end{split}
\end{equation}
As we shall see in the following section, the dynamics of the phase transition depend mainly on $\sigma$ and $\lambda$, while the strength of the gravitational wave signal depends on the details of the inflationary model and on the non-minimal coupling to gravity, i.e on $\beta_w$ and $\xi$. 

\begin{figure}[t]
    \centering
    \includegraphics[scale=0.99]{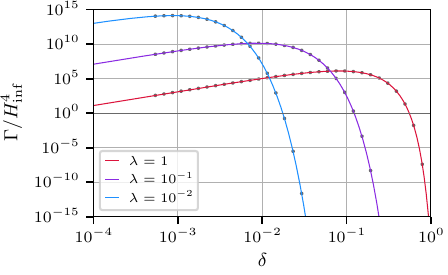}
    \includegraphics[scale=0.99]{{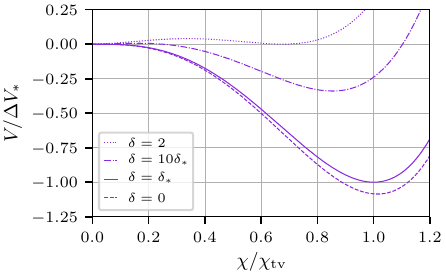}}
    \caption{\it{Left panel: Nucleation rate normalised to Hubble volume at the end of inflation for $\sigma/ H_{\rm inf}=100$ and different values of $\lambda$. The grey points represent the numerical values obtained from the computation of the bubble profiles, while the colourful lines are the fits from the analytical approximation \eqref{action}. Right panel: Temporal evolution of the scalar potential for the benchmark values $\sigma/ H_{{\rm inf}}=100$, $\lambda = 0.1$, and $\delta_{\ast} = 0.12$.}}
    \label{fig:Gamma}
\end{figure}

The probability of quantum tunneling in vacuum is typically measured via the bubble nucleation rate ~\cite{Fairbairn_2019, PhysRevD.46.2384}
\begin{equation}\label{eq:Gamma_def}
    \Gamma = \chi_{\textrm{tv}}^4\left(\frac{S_E}{2\pi}\right)^2 \exp(-S_E) \,, 
\end{equation}
where $\chi_{\textrm{tv}}$ is the vacuum expectation value that $\chi$ acquires after tunnelling and $S_E$ represents the Euclidean action associated with the $O(4)$-symmetric bounce solution
\begin{equation}
    S_E = 2\pi^2\int_0^{\infty} \varrho^3\textrm{d}\varrho\left[\left(\frac{1}{2}\frac{\textrm{d}\chi}{\textrm{d}\varrho} \right)^2 + V\right] \,.
\end{equation}
We avoid using well-known expressions that contain the radius of the critical bubble \cite{Coleman:1977py,Callan:1977pt,Linde:1980tt,Linde:1981zj, Ellis:2018mja} as the initial radius is not well defined in the limit of vanishing barrier, where the assumption about the thin-wall profile is not valid. Using an analytical formula derived from a numerical fit \cite{Adams:1993zs, Ellis:2020awk}, the Euclidean action of the critical bubble can be expressed as
\begin{equation}
    S_E = \frac{4\pi^2}{3\lambda} \frac{\alpha_1\delta + \alpha_2 \delta^2 + \alpha_3 \delta^3}{(2-\delta)^{3}} \,,
    \label{action}
\end{equation}
with $\alpha_1 = 13.832$, $\alpha_2 = -10.819$, $\alpha_3 = 2.0765$, and $\delta$ the potential-dependent coefficient defined in \eqref{eq:delta}. We have numerically verified the validity of this approximation at the limit of the vanishing barrier. The nucleation time $t_n$ is defined by the condition that, on average, one bubble per horizon is nucleated \cite{Ellis:2018mja}, namely
\begin{equation}
\int_{t_c}^{t_n} dt \frac{\Gamma(t)}{H(t)^3} = 1 \, ,
\label{eq:nucleation_time}
\end{equation}
where $t_c$ represents the moment when the two minima are degenerate. Assuming a constant Hubble rate during nucleation, the condition (\ref{eq:nucleation_time}) reduces to $\Gamma(t_n)=H(t_n)^4$, which will be used in this work. The moment $t_*$ when the transition completes is usually estimated as the moment of percolation~\cite{Ellis:2018mja, Ellis_2019, Kierkla:2022odc, Athron:2022mmm}. However, for relatively fast transitions $t_* \simeq t_n$ is a good approximation, which we adopt for simplicity. The evolution of the decay rate and the scalar potential for benchmark values of the couplings are shown in Figure~\ref{fig:Gamma}.

An important parameter of the phase transition is its strength $\alpha$, which scales according to the amount of vacuum energy released with respect to the energy of the cosmological background~\cite{LISACosmologyWorkingGroup:2022jok, Ellis:2020awk}. During the period of kination, the strength of the transition is given by
\begin{equation}
\label{eq:alpha-approx}
    \alpha \equiv \frac{\rho_V}{\rho_{\rm kin}} = \frac{\Delta V}{3 M_P^2 H_{\rm inf}^2} \approx \frac{\sigma^4 \left[ 1 + \mathcal{O} (\delta)\right]}{36 \lambda^3 M_P^2 H^2_{\rm inf}} \, .
\end{equation}
Another crucial parameter for describing the phase transition, and the resulting GW spectrum, is its time or length scale, which is associated with the inverse duration $\beta$ of the transition,
\begin{align}
    \beta = -\frac{d}{dt}S_E\Big{|}_{t=t_n} \, .
    \label{eq:beta_def}
\end{align}
In order to derive the value of the $\beta$ parameter, one needs to specify the actual realisation of the temporal dependence of the $\delta$ parameter. We define $\xi_{\textrm{min}}$ in such a way that it is the smallest possible non-minimal coupling ensuring that during inflation the true vacuum is located at $\phi=0$, and the secondary minimum, if present, is at most degenerate with the former
\begin{equation}\label{eq:ximin}
    \xi\geq \xi_{\textrm{min}} = \frac{\sigma^2}{54\lambda H_{\rm inf}^2}\,.
\end{equation}
Taking into account that the time-dependent mass is given by $M(t) = \xi \mathcal{R}(t)$ together with the parameterisation for the e.o.s.~in Eq.~\eqref{eq:omega}, we can derive an analytic expression for $\beta_*$ in terms of the value of $\delta$ at the time of nucleation (see Figure \ref{fig:delta_n}),
\begin{equation}\label{eq:beta-FULL}
\begin{split}
    \beta_* &= \frac{ 108 \pi^2 \beta_w  H_*^2  \xi  }{\sigma ^2\cosh^2({\beta_w  (t_*-t_0))} (2-\delta_* )^3} \left[ \left( \alpha_1+ 2 \alpha_2 \delta_* +3 \alpha_3 \delta_* ^2\right)+ \frac{3 \left(\alpha_1 \delta_* +\alpha_2 \delta_* ^2+\alpha_3 \delta_* ^3\right)}{2-\delta_* }\right] =\\
    &=\frac{{108 \pi^2 \beta_w H_*^2 (2 \alpha_1 (1 + \delta_*) + \delta_* (6 \alpha_3 \delta_* + \alpha_2 (4 + \delta_*))) \xi \left(1 - \left(\frac{1}{3} - \frac{\delta_* \sigma^2}{81 \lambda \xi H_{\ast}^2}\right)^2\right)}}{{(-2 + \delta_*)^4 \sigma^2}}\geq\\   
    &\geq\frac{8 \pi ^2 \beta_w (\delta_* +1) (2 \alpha_1 (\delta_* +1)+\delta_*  (\alpha_2 (\delta_* +4)+6 \alpha_3 \delta_* ))}{9 (2-\delta_* )^3 \lambda }\,,
\end{split}
\end{equation}
where we have removed the explicit time dependence in the second line by inverting the relation for $t(M)$ from Eqns. (\ref{eq:Ricci}) and (\ref{eq:omega}),
\begin{equation}
    \label{eq:nucleation-time}
     t_* - t_0 = \beta_w^{-1}{\rm arctanh} \left(\frac{1}{3}-\frac{\delta_{\ast} \sigma^2}{81 \lambda \xi H^2_*1}\right),
\end{equation}
and using hyperbolic trigonometric relations. The lower bound in the third line is obtained by fixing $\xi$ to the minimal value from Eq. \eqref{eq:ximin}.
The only viable source of gravitational wave signal generated during the non-thermal phase transitions are collisions between true vacuum bubbles. The spectrum of the signal produced at the time of the transition is described by \cite{Allahverdi:2020bys}
\begin{equation} \label{Omega_star}
    \Omega_* (f) = 
    \left(\frac{\beta_*}{H_*}\right)^{-2} 
    \qty(\frac{\rho_{ V}}{\rho_{\rm total}})^2
    S(f)\,,
\end{equation}
where the spectral shape $S(f)$ is defined using a numerically-derived broken power law~\cite{Lewicki:2020azd, Lewicki:2019gmv}
\begin{equation} \label{eq:S}
S\qty(f) = 25.10 \, \qty[ 2.41 \, \qty(\frac{f}{f_*})^{-0.56} + 2.34 \, \qty(\frac{f}{f_*})^{0.57} ]^{-4.2}\,, 
\end{equation}
with  $f_* = 0.13 \beta_*$ defined as the peak frequency at the time of production. The energy budget factor in Eq.~\eqref{Omega_star} can be calculated using the transition strength parameter $\alpha$ (see Eq.\eqref{eq:alpha-approx}),
\begin{align}
    \frac{\rho_{ V}}{\rho_{\rm total}} = 
    \frac{\rho_V}{\rho_V + \rho_{\rm kin}} = \frac{\alpha}{\alpha+1} \,.
\end{align}

Then, in order to obtain the present-day spectrum we have to redshift the signal, i.e. re-scale the amplitude and peak frequency. This must be done with caution since the transition takes place during the kination domination period. Splitting the redshift factor into a kination part and the regular radiation domination contribution results in \cite{Allahverdi:2020bys}
\begin{align}
\label{eq:GWredshift1}
    \Omega_{\rm GW, 0} &= \qty(\frac{a_*}{a_0})^4 \qty(\frac{H_*}{H_0})^2 \Omega_{*} =
    \qty(\frac{a_*}{a_{\rm RD}})^4 \qty(\frac{a_{\rm RD}}{a_0})^4
    \qty(\frac{H_*}{H_{\rm RD} })^2 \qty(\frac{H_{\rm RD}}{H_0})^2 \Omega_{*} \nonumber \\
    &= 
    \frac{1.67 \cross 10^{-5} \mbox{ Hz }}{h^2} 
    \qty(\frac{\alpha}{\alpha+1})^2
    \qty(\frac{H_*}{H_{\rm RD}})^{2\frac{3w-1}{3w+3}}
    \left(\frac{\beta_*}{H_*}\right)^{-2} S(f) \,, \\
    %%%%%%%%
    %%%%%%%%
    f_{\rm peak, 0} &= \frac{a_*}{a_0}f_* = 
    \frac{a_{\rm RD} H_{\rm RD}}{a_0} \frac{a_* H_*}{a_{\rm RD} H_{\rm RD}} \frac{f_*}{H_*} \nonumber\\
    &= 
    1.65 \cross 10^{-5} \mbox{ Hz } 
    \qty(\frac{T_{\rm RD}}{100 \mbox{ GeV}})
    \qty(\frac{f_*}{H_*})
    \qty(\frac{H_*}{H_{\rm RD}})^{\frac{3w-1}{3w+3}}\,.
\end{align}
The temperature at the onset of radiation domination can be calculated straight from the Hubble parameter, under the assumption of instantaneous thermalisation, as
\begin{align}
    T_{RD} = \qty(3 M_{p}^2 \xi_g ^2 H_{\rm RD}^2)^{\frac 14} \,,
\end{align}
where $\xi_g = \sqrt{30/\pi^2 g_*}$, and $1\leq g_* \lesssim 106.75$ is the number of effective degrees of freedom at $t_{\rm RD}$. As we do not specify the decay rate of the spectator field here, we will use the upper limit coming from the SM, assuming that $\chi$ has already decayed into the SM plasma before radiation domination. Different scenarios can also be considered here, but the overall impact on the spectra will not be significant. Finally, one has to take into account the implications of the modified expansion for the spectral shape. In particular, for kination, the slope of the GW spectrum changes to $\Omega_{\mathrm{GW}}(f) \propto f^4$~\cite{Gouttenoire:2021jhk} for modes entering the horizon before going back to the standard $\Omega_{\mathrm{GW}}(f) \propto f^3$ \cite{Durrer_2003,PhysRevD.79.083519,Cai_2020} as the radiation domination period begins, i.e.
\begin{align}
    \Omega_{\mathrm{GW}}(f) \propto 
    \begin{cases}
        S(f)  & \quad\text { for } f \gtrsim  \frac{a_*}{a_0} \frac{H_*}{2\pi}\,,\\
        f^4 & \quad \text { for } f \lesssim  \frac{a_*}{a_0} \frac{H_*}{2\pi}\,, \\ 
    f^3 & \quad\text { for } f \lesssim \frac{a_{\rm RD}}{a_0} \frac{H_{\rm RD}}{2\pi}\,,
    \end{cases}
    \label{eq:tilt}
\end{align}
with $S(f)$ denoting the model-dependent spectral shape defined in Eq.~\eqref{eq:S}. Notice that the factors present in the comoving Hubble radius $\frac{a_*}{a_0} H_*$ can be easily computed by using Eq.~\eqref{eq:GWredshift1}. 
Moreover, the value of the parameter $\alpha$ inevitably poses an upper bound on the duration of the kination epoch. Since
\begin{align}
    \eval{ \frac{\rho_{\rm rad}}{\rho_{\rm kin}} }_{a=a_{\rm RD}} = 
    \left(\frac{\rho_{\rm rad,0}}{\rho_{\rm kin}} \right) a_{\rm RD} ^2 =
    \alpha a_{\rm RD} ^2 = 1\,,
\end{align}
we are able to express the ratio of Hubble parameters simply as
\begin{align}
    \qty(\frac{H_*}{H_{\rm RD}})^2 = 
    \frac{\rho_{\rm kin}}{\rho_{\rm rad,RD}} = 
    \frac{\rho_{\rm kin}}{\rho_{\rm rad,0} a^{-4}_{\rm RD}} = 
    \alpha ^{-3}\,.
\end{align}
Now, combining these expressions with Eq.~\eqref{eq:GWredshift1} and setting $w=1$, we can obtain the following simple formulae for the current GW spectra and frequency,
\begin{align}
%%Omega
    \Omega_{\rm GW,0} &= 
    \frac{1.67 \cross 10^{-5}}{h^2} 
%    \frac{1}{(\alpha+1)^2}
    \alpha
    \left(\frac{\beta_*}{H_*}\right)^{-2} S(f)\,, \\
%% frequency
    f_{\rm peak,0 } &= 
    1.65 \cross 10^{-5} \mbox{ Hz } 
    \frac{\sqrt[4]{3 M_p^2 \xi_g^2 \alpha H^2_*}}{100 \mbox{ GeV}}
    \qty(\frac{f_*}{H_*})\,,
\end{align}
where we have dropped the negligible $(\alpha+1)^{-2}$ factor because the inflaton background energy density dominates at the time of the transition.

%%%%%%%%%%%%%%%%%%%%%%%%%%%%%%%%%%%%%%%%%%%%%%%%%%%%%%%%%%%%%%%%%%%%%%%%
\section{Parameter space and gravitational wave signal}\label{sec:gw}
\begin{figure}[t]
    \centering
    \includegraphics[scale=1]{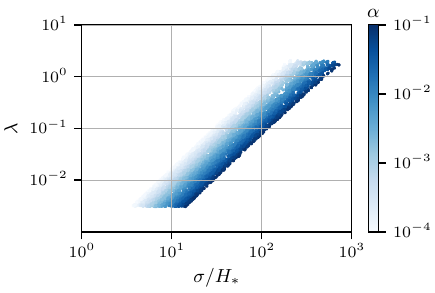}
    \includegraphics[scale=1]{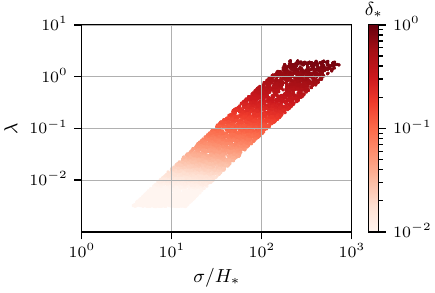}
    \includegraphics[scale=1]{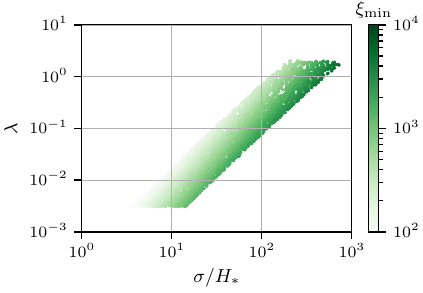}
    \includegraphics[scale=1]{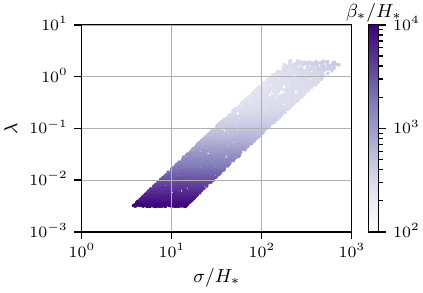}
    \caption{\it{Strength of the transition $\alpha$ (upper left panel), $\delta$ parameter at the moment of nucleation (upper right panel), lower bound on non-minimal coupling $\xi$ from Eq. \eqref{eq:ximin} (lower left panel) and duration of the transition $\beta_*/H_*$ evaluated for $\xi_{\textrm{min}}$ (lower right panel). The scans were performed for $H_{\ast} = 10^{12} \textrm{ GeV}$} and $\beta_w/H_{\rm inf} = 1$.} 
    \label{fig:delta_n}
\end{figure}

%%%%%%%%%%%%%%%%%%%%%%%%%%%%%%%%%%%%%%%%%%%%%%%%%%%%%%%%%%%%%%%%%%%%%%%%
We present the scan of the parameter space of our model, displaying plausible regions leading to first-order phase transitions with strengths compatible with the assumption of initial kination domination. We performed a scan for the discussed scalar potential and computed the relevant parameters for the phase transition, as shown in Fig. \ref{fig:delta_n}. 
We restrict ourselves to the interval $10^{-4}<\alpha<10^{-1}$. The lower bound comes from the fact that lower values of $\alpha$ lead to very weak transitions. The upper bound is simply a requirement ensuring that the energy density of the field remains subdominant during the evolution of the e.o.s.~parameter. In fact, if the difference $\Delta V$ between the false and the true vacuum reaches values comparable to the background energy density, we would need to consider its contribution to the total energy-momentum tensor. Since it scales like vacuum energy, it would tend to restore the e.o.s.~parameter to $w=-1$. Similarly, we have focused on the values for which $\beta_w/H<\beta_*/H_*<10^4$, where the hierarchy between the rate of the FOPT and the shift from inflation to kination is maintained. Higher values of $\beta_*$ would lead to transitions that closely resemble second-order ones and would not produce any visible signal.

\begin{figure}[!t]
    \centering
    \includegraphics[scale=1]{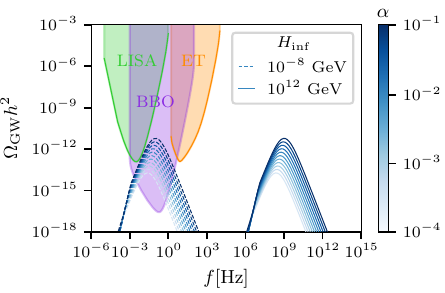}
    \includegraphics[scale=1]{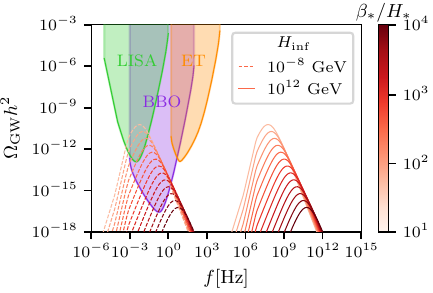}
    \caption{\it{Gravitational waves spectra from phase transitions triggered at the onset of kination. In the left panel, $\alpha$ is treated as a free parameter while $\beta_*/H_{\ast} = 100$. On the right panel $\alpha = 0.01$, while $\beta_*/H_{\ast}$ is treated as the free parameter. Solid lines represent typical results for the quintessential inflation ($H_{\rm inf}\approx 10^{12}$ GeV), dashed ones the case of ultra-low inflationary scale ($H_{\rm inf}\approx 10^{-8}$ GeV). The shaded, colourful regions denote the sensitivity ranges of future experiments. }}
    \label{fig:gw_spectra_alpha_beta}
\end{figure} 

Regarding the computation of the GW spectra, we consider a typical inflationary scale of ${H_{\rm inf} \sim 10^{12} \text{ GeV}}$, and a lower scale, ${H_{\rm inf} \sim 10^{-8} \text{ GeV}}$ as shown in Figure~\ref{fig:gw_spectra_alpha_beta}. 
The change of the transition strength $\alpha$ for constant inverse time scale $\beta_*/H_{\ast}$ trivially affects the amplitude. However, due to the non-standard redshift scenario, larger values of $\alpha$ translate generically into higher frequencies. Varying the timescale $\beta_*/H_{\ast}$ of transition produces the usual effect: the smaller value it has, the longer the transition is or the ``bigger'' the bubbles are, which results in larger amplitudes of the signal and lower peak frequencies. Moreover, in both cases one can observe the ``drop'' in the slope for the super-horizon modes as described in Eq. \eqref{eq:tilt}. This effect is important, since it not only affects detection prospects, but also because finding this feature in the observed GW spectrum would be evidence of a kination period in the early Universe.

\begin{figure}[!h]
    \centering
    \includegraphics[scale=1]{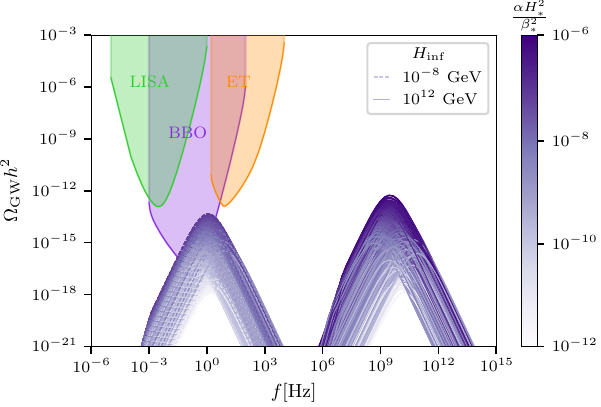}
    \caption{\it{
    Gravitational waves spectra from phase transition triggered by onset of kination period. The signal comes from the actual points from the scan of parameter space (see Fig. \ref{fig:delta_n}). Solid lines represent typical results for the quintessential inflation (${H_{\rm inf}=  10^{12}}$ {\rm GeV}), dotted ones the case of ultra-low inflationary scale (${H_{\rm inf} = 10^{-8}}$ {\rm GeV}). The shaded, colourful regions denote sensitivity range of future experiments.
    }}
    \label{fig:gw_spectra_scan}
\end{figure}

Figure~\ref{fig:gw_spectra_scan} shows the spectra resulting from the scans we performed. Note that the signal from the transitions at the inflationary scales ${H_* \sim 10^8-10^{12} \text{ GeV}}$ is not in the reach of existing and future experiments.  However, the signal can be within reach of future detectors such as the Einstein Telescope (ET) \cite{Punturo:2010zz}, the Big Bang Observer (BBO) \cite{Harry:2006fi} or even the Laser Interferometer Space Antenna (LISA) \cite{LISA} for more exotic inflationary scenarios taking place closer to EW scale.

%%%%%%%%%%%%%%%%%%%%%%%%%%%%%%%%%%%%%%%%%%%%%%%%%%%%%%%%%%%%%%%%%%%%%%%%
\section{Conclusions} \label{sec:conclusions}
%%%%%%%%%%%%%%%%%%%%%%%%%%%%%%%%%%%%%%%%%%%%%%%%%%%%%%%%%%%%%%%%%%%%%%%%
The advent of the gravitational wave cosmology era presents us with an unprecedented possibility to explore the pre-CMB epoch, allowing us to perform a systematic search for new physics throughout energy scales. Both in the context of SM and BSM physics, first-order phase transitions embody the prototypical phenomenon that can produce a sizeable stochastic gravitational wave background that could be detected by future experiments. Given the broad range of possible frequencies and different modalities within which such transitions could take place, it is of the utmost importance to explore non-standard scenarios beyond those usually embedded in the SM, such as transitions at the EW or the QCD scale.

In this work, we analyzed the possibility of having a FOPT triggered by the change of the space-time geometry, contrary to standard scenarios where the clock for the transition is set by the thermal evolution of the Universe. In our case, the transition proceeds through quantum tunnelling to a true vacuum that dynamically appears due to the evolution of the e.o.s. parameter, and therefore the scalar curvature. In this context, we showed that a spectator field with a renormalisable potential and a non-minimal coupling to gravity can successfully lead to a phase transition under quite general conditions. As a proof of concept, we focused on the post-inflationary evolution of the Universe in the context of quintessential inflation. This choice provides an ideal setup to explore all the values of the e.o.s. in the range $-1\leq w \leq 1$, while providing a subsequent enhancement of the signal during the kination period. Moreover, this kind of scenario naturally sets an upper bound on the duration of reheating, as the produced radiation will eventually overcome the kination background, without the requirement of specifying any coupling to the SM particles.

Our results show that the strength $\alpha$ of the transition depends only on the cubic and quartic terms of the potential, while its speed $\beta$ is also regulated by the details of the time evolution of the e.o.s. parameter. We presented the predictions for gravitational wave spectra resulting from vacuum bubble collisions, taking into account recent developments from the simulations and the non-standard redshift for superhorizon modes caused by the kination period. We found the relation between the released energy i.e. strength of the transition and the peak frequency, and estimated the detection prospects for different energy scales of inflation. As expected for a typical inflationary scale $H_{\rm inf}\sim 10^{12}$ GeV, the frequency of the signal lies outside of the observable range covered by current experiments. Moreover, we find that a signal compatible with LISA and BBO sensitivities would require an inflationary scale of the order $H_{\rm inf} \sim 10^{-8}$ GeV, which is typically incompatible with quintessential inflationary scenarios.

Our analysis constitutes a first step towards accurate predictions for expansion-driven phase transitions. However, there are multiple improvements that can be made in future studies. In our approach we assumed that the flat space-time limit of the field mass $m$ is negligible, leading to dynamics solely determined by the non-minimal coupling and the $\beta_w$ parameter. In the presence of a non-negligible value of $m$, one might tune the parameters in order to delay the nucleation condition very close to the onset of kination. In such a case, due to the chosen parameterisation of the e.o.s., lower values of $\beta_*$ can be achieved. The framework presented in this paper can also be adjusted to study different realizations of the scalar potential with various field content and gravitational couplings. One can also analyze various inflationary models, going beyond the quintessential inflation, in which case the methods described in this work allow for a novel probe of the early Universe.

%%%%%%%%%%%%%%%%%%%%%%%%%%%%%%%%%%%%%%%%%%%%%%%%%%%%%%%%%%%%%%%%%%%%%%%%%%%%%%%%
\acknowledgments
%%%%%%%%%%%%%%%%%%%%%%%%%%%%%%%%%%%%%%%%%%%%%%f%%%%%%%%%%%%%%%%%%%%%%%%%%%%%%%%%%
We thank Dr.~Michał Artymowski for useful discussions. % and drinks  
We acknowledge funding from the Polish National Agency for Academic Exchange (NAWA) and the Fundação para a Ciência e a Tecnologia (FCT) within the bilateral Programme for Cooperation in Science between Portugal and Poland, project 2021.09261.CBM. 
This work was supported by the Polish National Agency for Academic Exchange within Polish Returns Programme under agreement PPN/PPO/2020/1/00013/U/00001 and the Polish National Science Center grant 2018/31/D/ST2/02048.
J.~R. is supported by a Ram\'on y Cajal contract of the Spanish Ministry of Science and Innovation with Ref.~RYC2020-028870-I.  
G.~L. is supported by a fellowship from ”la Caixa” Foundation (ID 100010434) with fellowship code LCF/BQ/DI21/11860024.
G.~L. and M.~P. acknowledge the Funda\c c\~ao para a Ci\^encia e a Tecnologia (FCT), Portugal, for the financial support to the Center for Astrophysics and Gravitation-CENTRA, Instituto Superior T\'ecnico,  Universidade de Lisboa, through the Project No.~UIDB/00099/2020.  M.~P. thanks  also the support of this agency through the Grant No. SFRH/BD/151003/2021 in the framework of the Doctoral Program IDPASC-Portugal.

%%%%%%%%%%%%%%%%%%%%%%%%%%%%%%%%%%%%%%%%%%%%%%%%%%%%%%%%%%%%%%%%%%%%%%%%

%%%%%%%%%%%%%%%%%%%%%%%%%%%%%%%%%%%%%%%%%%%%%%%%%%%%%%%%%%%%%%%%%%%%%%%%
\appendix

\bibliography{main}

\end{document}